\documentclass{svproc}

\usepackage[version=3]{mhchem}
\usepackage[printonlyused, nolist]{acronym}
\usepackage{braket}
\usepackage{graphicx}
\usepackage{dcolumn}
\usepackage{bm}
\usepackage{subcaption}
\usepackage{placeins}
\usepackage{url}
\usepackage{amsmath,amssymb}
\usepackage{cleveref}
\usepackage{graphicx}
\usepackage{subcaption}
\usepackage{pgf}
\usepackage{tikz}
\usepackage{xcolor}
\usetikzlibrary{arrows,positioning,shapes,fit,backgrounds,decorations,scopes,calc}

\begin{acronym}
  \acro{ml}[ML]{Machine Learning}
  \acro{qml}[QML]{Quantum Machine Learning}
  \acro{hqcml}[HQCML]{Hybrid Quantum--Classical Machine Learning}
  \acro{vqa}[VQA]{Variational Quantum Algorithm}
  \acroplural{vqa}[VQAs]{Variational Quantum Algorithms}

  \acro{pca}[PCA]{Principal Component Analysis}
  \acro{pls}[PLS]{Partial Least Squares}
  \acro{plsr}[PLSR]{Partial Least Squares Regression}

  \acro{fast}[Quantum-FAST]{Minimal hybrid model (quantum-first analysis of self-training)}
  \acro{hybridplus}[HybridPlus]{Refined hybrid model (entanglement- and dimension-adjusted)}

  \acro{qiskit}[Qiskit]{Open-source framework for quantum computing}
  \acro{qiskitml}[Qiskit ML]{Qiskit Machine Learning}
  \acro{qnn}[QNN]{Quantum Neural Network}
  \acro{estqnn}[EstimatorQNN]{Estimator-based Quantum Neural Network}
  \acro{pytorch}[PyTorch]{PyTorch deep learning framework}
  \acro{sklearn}[scikit-learn]{\texttt{scikit-learn} machine learning library}

  \acro{acc}[ACC]{Accuracy}
  \acro{ari}[ARI]{Adjusted Rand Index}
  \acro{nmi}[NMI]{Normalized Mutual Information}

  \acro{tsi}[TSI]{Training Stability Index}
  \acro{tei}[TEI]{Training Efficiency Index}
  \acro{qgn}[QGN]{Quantum Gradient Norm}
  \acro{bpi}[BPI]{Barren Plateau Indicator}

  \acro{edqfs}[EDQFS]{Effective Dimension of the Quantum Feature Space}
  \acro{qos}[QOS]{Quantum Output Sensitivity}
  \acro{fmcr}[FMCR]{Feature Map Compression Ratio}
  \acro{qlad}[QLAD]{Quantum Layer Activation Diversity}

  \acro{qce}[QCE]{Quantum Circuit Expressibility}
  \acro{qvc}[QVC]{Quantum Volume Contribution}
  \acro{qcf}[QCF]{Quantum Circuit Fidelity}
  \acro{qlr}[QLR]{Quantum Locality Ratio}
  \acro{eee}[EEE]{Effective Entanglement Entropy}
  \acro{qmi}[QMI]{Quantum Mutual Information}
\end{acronym}

\begin{document}
\mainmatter

\title{From Classical to Hybrid: A Practical Framework for Quantum-Enhanced Learning}

\titlerunning{A Practical Framework for Quantum-Enhanced Learning}

\author{Silvie Ill\'{e}sov\'{a}\inst{1} \and
Tom\'{a}\v{s} Bezd\v{e}k\inst{2} \and
Vojt\v{e}ch Nov\'{a}k \inst{3,4} \and
Ivan Zelinka\inst{3,5} \and
Stefano Cacciatore \inst{6} \and
Martin Beseda\inst{7}
}

\authorrunning{S.Ill\'{e}sov\'{a} et al.}

\tocauthor{Silvie Ill\'{e}sov\'{a}, Tom\'{a}\v{s} Bezd\v{e}k, Vojt\v{e}ch Nov\'{a}k, Ivan Zelinka, Stefano Cacciatore, Martin Beseda}

\institute{
Gran Sasso Science Institute, L’Aquila, Italy \and
Department of Mathematics, TUM School of Computation, Information and Technology, Technical University of Munich, Garching bei München, Germany \and
Department of Computer Science, Faculty of Electrical Engineering and Computer Science, VSB - Technical University of Ostrava, Ostrava, Czech Republic\and
IT4Innovations National Supercomputing Center,\\
VSB - Technical University of Ostrava, 708 00 Ostrava, Czech Republic \and
Department of Informatics and Statistics, Marine Research Institute,\\
Klaipeda University, Lithuania
\and
International Centre for Genetic Engineering and Biotechnology
Wernher and Beit Building (South)
Anzio Road, Observatory 7925
Cape Town, South Africa \and
Dipartimento di Ingegneria e Scienze dell’Informazione e Matematica,\\ 
Università dell’Aquila, Via Vetoio, I-67010 Coppito, L’Aquila, Italy 
}

\maketitle

\begin{abstract}
This work addresses the challenge of enabling practitioners without quantum expertise to transition from classical to hybrid quantum–classical machine learning workflows. We propose a three-stage framework: starting with a classical self-training model, then introducing a minimal hybrid quantum variant, and finally applying diagnostic feedback, via \texttt{QMetric} to refine the hybrid architecture. In experiments on the Iris dataset, the refined hybrid model improved accuracy from 0.31 classical to ~0.87 in quantum approach. These results suggest that even modest quantum components, when guided by proper diagnostics, can enhance class separation and representation capacity in hybrid learning, offering a practical pathway for classical Machine Learning practitioners to leverage quantum-enhanced methods
\keywords{Hybrid quantum-classical learning, Quantum machine learning, Variational quantum circuits, Training diagnostics, Feature representation}

\end{abstract}

\section{Introduction}
In the current state of the art, classical machine learning models, workflows, and overall approaches are well documented and widely accessible \cite{zhou2021machine,jordan2015machine,pandey2019machine}. However, with recent developments in quantum computing, a new approach has emerged, namely quantum machine learning \cite{biamonte2017quantum,cerezo2022challenges}, or hybrid quantum–classical machine learning \cite{pulicharla2023hybrid,liu2021hybrid}. This provides potential advantages in representation capacity \cite{banchi2021generalization,illesova2025importance} and, in the future, may enable faster training processes \cite{schuld2022quantum,mishra2020quantum}.
Currently the application of quantum computing is wide, ranging from chemistry applications \cite{cao2019quantum,rajamani2025equiensembledescriptionsystematicallyoutperforms,illesova2025numerical,illesova2025transformation}, physics advancements \cite{di2024quantum,micheletti2021polymer,PhysRevA.111.022437,lamm2020parton,mocz2021toward}, benchmarking applications \cite{lewandowska2025benchmarking,bilek2025experimental,proctor2025benchmarking,evolvingcircuits,lubinski2023application,hashim2025practical,illesova2025qmetric}, material design calculations \cite{guo2024harnessing,de2021materials,kang2025quantum,liu20192d,ma2020quantum}, the field of software engineering  \cite{trovato2025preliminary,piattini2021quantum,ali2022software,dwivedi2024quantum}, and the finance sector \cite{yuan2024quantifying,egger2020quantum,rebentrost2018quantum,herman2023quantum}. And while many, approaches specialized in qunatum machine learning exists \cite{gupta2022quantum,illesova2025importance,novak2025predicting,biamonte2017quantum,aggarwal2024detailed}, the practical adoption of these approaches is often difficult for those who do not have experience with quantum computing and quantum information.

The main challenge is the lack of a clear workflow for transitioning from classical models to hybrid quantum models without requiring direct quantum expertise. To address this, we propose a practical workflow.

The starting point is a classical machine learning model. Next, a minimal hybrid model with the same learning structure is designed. This model is analyzed using a toolkit to assess training behavior, feature-space properties, and quantum circuit characteristics; in our case, the toolkit of choice is \texttt{QMetric} \cite{illesova2025qmetric}. We analyze the feedback and apply targeted modifications to the minimal model to improve its training characteristics, based on its identified weak points. This refinement can be performed iteratively until the desired hybrid model is obtained.

We demonstrate this workflow using an intentionally simplified model on the Iris dataset \cite{fisher1936use}, together with a self-training label refinement procedure, which is implemented in \texttt{Qiskit} \cite{javadi2024quantum}, \texttt{Qiskit Machine Learning} \cite{sahin2025qiskit}, \texttt{PyTorch} \cite{paszke2019pytorch} and \texttt{scikit-learn} \cite{pedregosa2011scikit}. We compare three models, the classical model, the minimal hybrid quantum model, and the improved hybrid model.

The main contribution of this work is a repeatable and concrete workflow that enables classical machine learning practitioners to benefit from quantum resources without requiring quantum expertise.

\section{Methodology}

In this section, the three models, that are used in the proposed framework are described, with specifications, and highlighted differences.

\subsection*{Classical Baseline}

The chosen simplified model, that was chosen as the classical starting point for the demonstrated approach is a a self-training model based on Partial Least Squares Regression \cite{geladi1986partial,tobias1995introduction}. The chosen input is a subset of the Iris dataset, which si represented as a matrix
 $X \in \mathbb{R}^{N \times d}$ with $N = 150$ samples and the number of features is $d = 4$.  The initial label assignment is arbitrarily generated, as each sample is given its index as an starting label. In this way, the starting label space is effectively maximally unstructured label space. The goal of the procedure is to progressively refine these labels toward a stable clustering structure.

At iteration $t$, let $y^{(t)}$ denote the current label estimates. These labels are binarized and used to fit a Partial Least Squares Regression model, which produces predicted labels $\hat{y}^{(t)}$ using cross-validation. The self-training update rule is then

\begin{equation}
y^{(t+1)}_i =
    \begin{cases}
\hat{y}^{(t)}_i, & \text{if } \hat{y}^{(t)}_i \text{ is consistent and improves classification accuracy}, \\
y^{(t)}_i,       & \text{otherwise.}
\end{cases}
\end{equation}

These label updates are applied only to a subset of samples that the model misclassifies under cross-validation procedure. This process causes the changes in labels to be gradual, and only some mismatched labels are replaced with the model’s predicted labels at each iteration.

This iterative process in theory continues until no further accuracy improvement is observed, but in our case we impose an arbitrary iteration limit of $n_{maxit} = 20$. The resulting label assignments form the intentionally not well trained baseline, so that the work on hybrid model may start.

\subsection{Minimal Hybrid Model}

The first hybrid model, referred to as \texttt{Quantum-FAST}, is constructed to closely mirror the structure of the classical baseline while introducing a minimal quantum component. The input data is first reduced to four principal components using Principal Component Analysis \cite{abdi2010principal}, ensuring a compact feature representation suitable for encoding into a quantum circuit. The quantum component consists of a two qubit \texttt{EstimatorQNN} \footnote{\url{https://qiskit-community.github.io/qiskit-machine-learning/stubs/qiskit_machine_learning.neural_networks.EstimatorQNN.html}}, followed by a small classical neural network head that maps the quantum outputs to label estimates. The forward mapping of the hybrid model can be written as
\begin{equation}
    \hat{y} = g_{\phi}\!\big(f_{\theta}(x)\big),
\end{equation}
where $f_{\theta}$ denotes the parameterized quantum circuit and $g_{\phi}$ denotes the classical neural head. The training loop, including the iterative self-training refinement procedure, is kept identical to the classical baseline to ensure comparability. Early stopping and gradient clipping \cite{liu2022communication} are applied to maintain training stability and to avoid divergence during optimization. This model serves as the initial hybrid reference point prior to any \texttt{QMetric} guided refinement.

\subsection{Refined Hybrid Model}

The second hybrid model, referred to as \texttt{HybridPlus}, is obtained by refining the initial \texttt{Quantum-FAST} model based on diagnostic feedback provided by \texttt{QMetric}. The analysis of the minimal hybrid model indicated limited entanglement and restricted feature diversity in the quantum embedding. To address these issues, an additional classical adapter layer was introduced before the quantum circuit, and the quantum ansatz was replaced with a deeper, fully entangling parameterized circuit. The resulting forward mapping takes the form
\begin{equation}
    \hat{y} = g_{\phi}\!\big(f_{\theta}(h_{\psi}(x))\big),
\end{equation}
where $h_{\psi}$ denotes the classical adapter layer, $f_{\theta}$ the refined quantum circuit, and $g_{\phi}$ the classical neural head. The indicators, which drove these changes, will be discussed in the \Cref{sec:results}.

The training loop, loss function structure, and iterative self-training refinement procedure remain unchanged to ensure a fair comparison with both the classical baseline and the minimal hybrid model. This refined model serves to evaluate the effect of \texttt{QMetric} guided architectural adjustments on model performance and training behavior.

\section{Results}\label{sec:results}

\begin{figure}[t]
  \centering
  \begin{subfigure}[t]{0.48\linewidth}
    \centering
    \includegraphics[width=\linewidth]{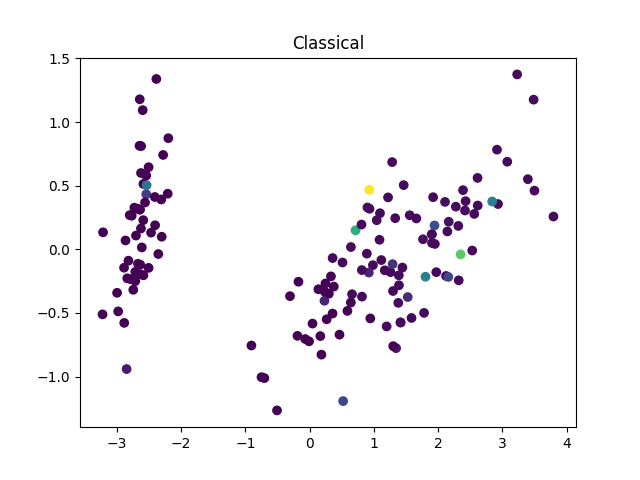}
    \caption{Classical baseline}
    \label{fig:clas}
  \end{subfigure}\hfill
  \begin{subfigure}[t]{0.48\linewidth}
    \centering
    \includegraphics[width=\linewidth]{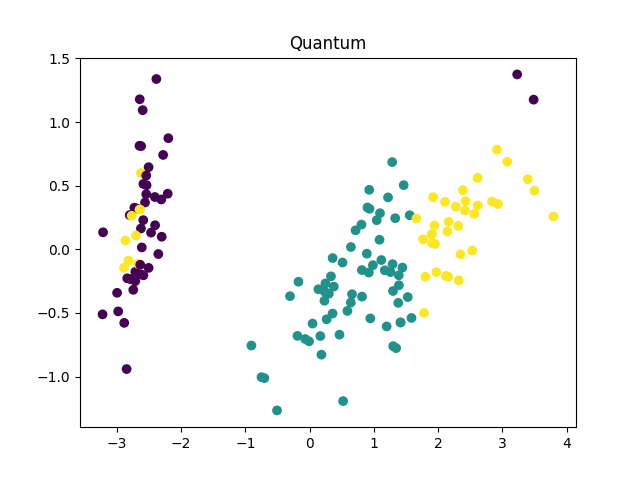}
    \caption{\texttt{Quantum-FAST}}
    \label{fig:fast}
  \end{subfigure}

  \vspace{0.6em}

  \begin{subfigure}[t]{0.48\linewidth}
    \centering
    \includegraphics[width=\linewidth]{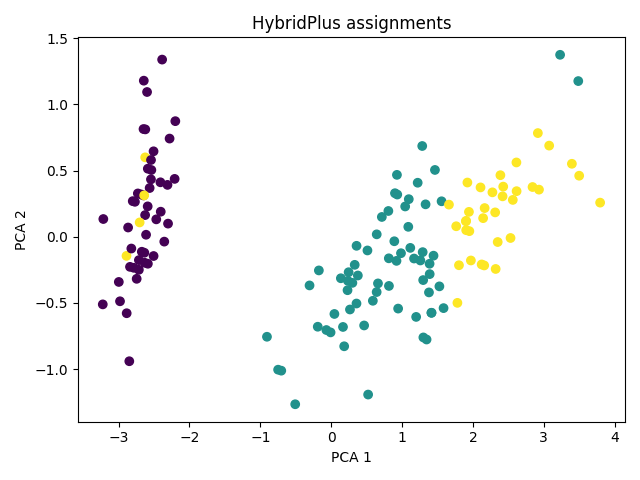}
    \caption{\texttt{HybridPlus}}
    \label{fig:plus}
  \end{subfigure}\hfill
  \begin{subfigure}[t]{0.48\linewidth}
    \centering
    \includegraphics[width=\linewidth]{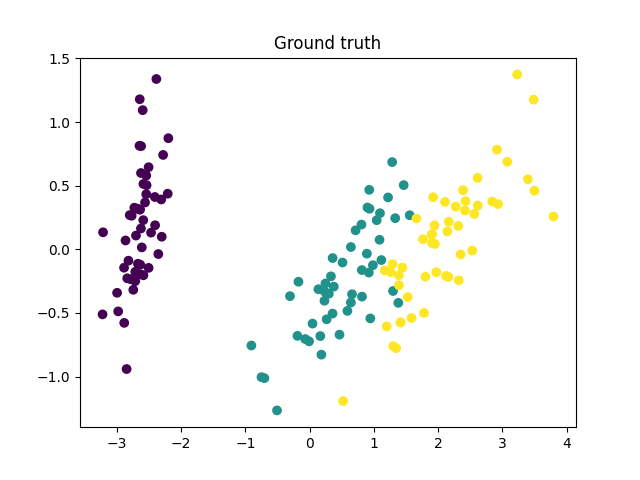}
    \caption{Ground truth}
    \label{fig:truth}
  \end{subfigure}

  \caption{Scatter plots in the learned feature space. (a) classical baseline, (b) \texttt{Quantum-FAST}, (c) \texttt{HybridPlus}, and (d) ground truth labels. 
  The quantum variants produce clearer class separation than the classical method, with \texttt{HybridPlus} aligning most closely with the true structure.}
  \label{fig:results}
\end{figure}

During the example calculations, the simplified classical self-labeling baseline model achieved, relatively high internal self-consistency score equal to $a_{internal} = 0.8333$, but the agreement to the true labels of the Iris dataset was weak as shown in \Cref{fig:clas}, while the true labels are depicted in \Cref{fig:truth}. However, when the results were compared to the true labels obtained from the dataset, this approach reached the accuracy of $A = 0.3067$, with the \ac{ari} \cite{santos2009use}, which quantifies how similar the clustering structure is to the true classes after correcting for chance, of $r = 0.2989$. The \ac{nmi} \cite{estevez2009normalized}, which measures how much information the cluster assignments share with the true labels, was equal to just $m = 0.3819$. These values indicate, that although the iterative relabeling process was stabilized, the final labels did not capture the geometry of the Iris dataset.

The \texttt{Quantum-FAST} model, which was used as the minimal hybrid model approach, achieved similar level of internal accuracy, $a_{internal} = 0.84$, as the classical model, but the number of agreements with the true labels was significantly higher, with accuracy of $a=0.8333$, \ac{ari} of $r=0.5928$ and \ac{nmi} of $m = 0.6183$. This improvement showcases one the strengths of hybrid approach, as the separation of data classes is better in the feature space, after encoding the data for quantum computers. The resulting labels are depicted in \Cref{fig:fast}

Now, to gain better insights, into how the representational structure looks, and also into the training dynamics, we utilize \texttt{QMetric}'s diagnostic ability. The \ac{tsi} was stable, equal to $TSI = 1.0$, and both the \ac{qgn}, $QGN = 0.0403$, and \ac{bpi}, $BPI = 0.000136$ confirm that optimization proceeded without the problem of vanishing gradients. The \ac{edqfs}, $EDQFS = 1.2952$ and \ac{qos}, $QOS = 5.38$ indicate a moderately expressive embedding. Circuit-level metrics show low entanglement, with \ac{eee} equal to $EEE = 0.175$ and limited \ac{qmi}, $QMI = 0.350$. This means that the representation of the data remains mostly local in structure. 

These observations motivated a refinement of the model toward stronger class-separating structure. In particular, the low \ac{eee} and \ac{qmi} indicated insufficient entanglement, while the moderate \ac{edqfs} suggested that the feature space was more expressive than actively used. Therefore, the \texttt{HybridPlus} variant increased entanglement depth and concentrated the feature space representation, aiming to sharpen decision boundaries while preserving training stability. This adjustment was intended to better align the embedding with the intrinsic geometry of the Iris dataset. So, in the \texttt{HybridPlus} variant, we increased the entangling depth of the variational circuit, added an additional layer of parameterized two qubit interactions to raise entanglement (\ac{eee}, \ac{qmi}), and reduced the effective feature space dimensionality through weight regularization, thereby concentrating the embedding around the class separating directions.

The \texttt{HybridPlus} model achieved a lower internal self-consistency of $a_{internal} = 0.6933$, reflecting that it does not preserve earlier label assignments as strongly; however, its agreement with the true class structure improved further, reaching $A = 0.8667$, $r = 0.6675$, and $m = 0.7093$. This confirms that the increased entanglement and more concentrated feature-space embedding helped to sharpen the class boundaries. The QMetric diagnostics support this shift as the \ac{edqfs} decreased to $EDQFS = 1.0009$, indicating a more focused representation, while the \ac{qos} rose significantly to $QOS = 37.93$, signaling stronger decisiveness in label assignment. The metrics at the circuit level also changed, with entanglement becoming more global, as shown by $EEE = 0.718$ and $QMI = 1.436$. Importantly, training remained stable, $TSI = 1.0$, and gradients did not collapse, $QGN = 0.0258$, $BPI = 0.000042$, demonstrating that the improvements in expressiveness were achieved without compromising learnability. The final labeling of this approach is visible in \Cref{fig:plus}.

\section{Conclusions}

This study successfully illustrated a realistic way to take an individual from classical self-training models to the potential for hybrid quantum-classical models without any prior quantum knowledge. Beginning from a simplified classical baseline, we built a minimal hybrid model, retaining the training pipeline, that showed even a small quantum contribution could improve learning labels and true structure of the Iris dataset. We conducted \texttt{QMetric} diagnostics to better understand training stability, representational characteristics, and circuit-level entanglement of the hybrid model. These insights led to targeted improvements to the quantum circuit and model architecture that resulted in the \texttt{HybridPlus} variant which improves class separation and greater agreement with ground truth.

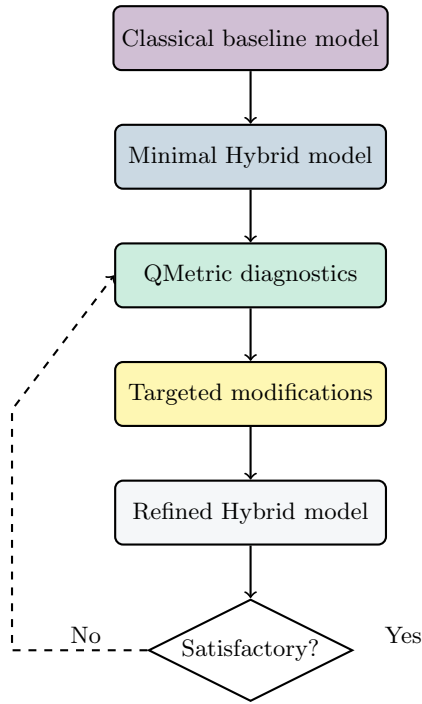
\begin{figure}[t]
\centering
\begin{tikzpicture}[
  node distance=7mm,
  box/.style={rectangle, rounded corners=3pt, draw=black, thick, text=black,
              minimum width=36mm, minimum height=8.5mm, align=center},
  decision/.style={diamond, draw=black, thick, fill=white, aspect=2,
                   align=center, inner sep=1.5pt},
  arrow/.style={->, thick}
]

\definecolor{viribase1}{RGB}{68,1,84}
\definecolor{viribase2}{RGB}{49,104,142}
\definecolor{viribase3}{RGB}{53,183,121}
\definecolor{viribase4}{RGB}{253,231,37}
\colorlet{pvir1}{viribase1!25!white}
\colorlet{pvir2}{viribase2!25!white}
\colorlet{pvir3}{viribase3!25!white}
\colorlet{pvir4}{viribase4!35!white}

\node[box, fill=pvir1] (clas) {Classical baseline model};
\node[box, fill=pvir2, below=of clas] (fast) {Minimal Hybrid model};
\node[box, fill=pvir3, below=of fast] (qmetric) {QMetric diagnostics};
\node[box, fill=pvir4, below=of qmetric] (modify) {Targeted modifications};
\node[box, fill=pvir2!20!white, below=of modify] (plus) {Refined Hybrid model};
\node[decision, below=of plus] (decide) {Satisfactory?};

\draw[arrow] (clas) -- (fast);
\draw[arrow] (fast) -- (qmetric);
\draw[arrow] (qmetric) -- (modify);
\draw[arrow] (modify) -- (plus);
\draw[arrow] (plus) -- (decide);

\draw[arrow, dashed]
  (decide.west) -- ++(-18mm,0) -- ++(0,32mm) -- (qmetric.west);

\node[right=3mm of decide, yshift=2mm] {\small Yes};
\node[left=5mm of decide,  yshift=2mm] {\small No};

\end{tikzpicture}
\caption{Proposed workflow with iterative refinement loop.}
\label{fig:workflow}
\end{figure}

The improvement was made without negatively impacting the stability of training or inducing barren plateau effects. Overall, the described work describes how hybrid models can be systematically improved through diagnostic feedback as opposed to trial-and-error model architecture. This demonstrates an achievable, replicable pathway for practitioners of classical machine learning to incorporate quantum computational elements into their models in a controlled, informed manner.

In summary, this work demonstrates that hybrid quantum models can be systematically improved through diagnostic feedback rather than intuition alone, showing not only how to refine the quantum feature space using tools such as \texttt{QMetric}. But also proposing that this refinement process can be repeated iteratively to progressively guide a classical model toward an increasingly effective hybrid design. The whole workflow is depicted in \Cref{fig:workflow}.

\section*{Acknowledgments}
Martin Beseda is supported by Italian Government (Ministero dell'Università e della Ricerca, PRIN 2022 PNRR) -- cod.P2022SELA7: ''RECHARGE: monitoRing, tEsting, and CHaracterization of performAnce Regressions`` -- Decreto Direttoriale n. 1205 del 28/7/2023. Vojtěch Novák is supported by Grant of SGS No. SP2025/072, VSB-Technical University of Ostrava, Czech Republic.

\section*{Data availability}
For reproducibility, all data, model configurations, and code used in this study are openly available in the associated GitLab repository at \url{https://gitlab.com/illesova.silvie.scholar/classical-to-hybrid}.
\appendix

\bibliographystyle{splncs04}
\bibliography{apssamp}

\end{document}